\documentclass[journal=jacsat,manuscript=article]{achemso}
\usepackage{chemformula} % Formula subscripts using \ch{}
\usepackage[T1]{fontenc} % Use modern font encodings
\usepackage{bm}
\usepackage{mathrsfs}
\usepackage{amssymb}
\usepackage{physics}
\usepackage{threeparttable}
\newlength{\LL} \LL 1\linewidth
\author{Almaz Khabibrakhmanov}
\affiliation{Department of Physics and Materials Science, University of Luxembourg, L-1511 Luxembourg City, Luxembourg}
\author{Dmitry V. Fedorov}
\affiliation{Department of Physics and Materials Science, University of Luxembourg, L-1511 Luxembourg City, Luxembourg}
\author{Alexandre Tkatchenko}
\affiliation{Department of Physics and Materials Science, University of Luxembourg, L-1511 Luxembourg City, Luxembourg}
\email{alexandre.tkatchenko@uni.lu}

\title[An \textsf{achemso} demo]
  {Universal Pairwise Interatomic van der Waals Potentials Based on Quantum Drude Oscillators}

\abbreviations{IR,NMR,UV}
\keywords{American Chemical Society, \LaTeX}

\begin{document}

\begin{tocentry}
%
%Some journals require a graphical entry for the Table of Contents. This should be laid out ``print ready'' so that the sizing of the text is correct.
%
%Inside the \texttt{tocentry} environment, the font used is Helvetica 8\,pt, as required by \emph{Journal of the American Chemical Society}.
\includegraphics{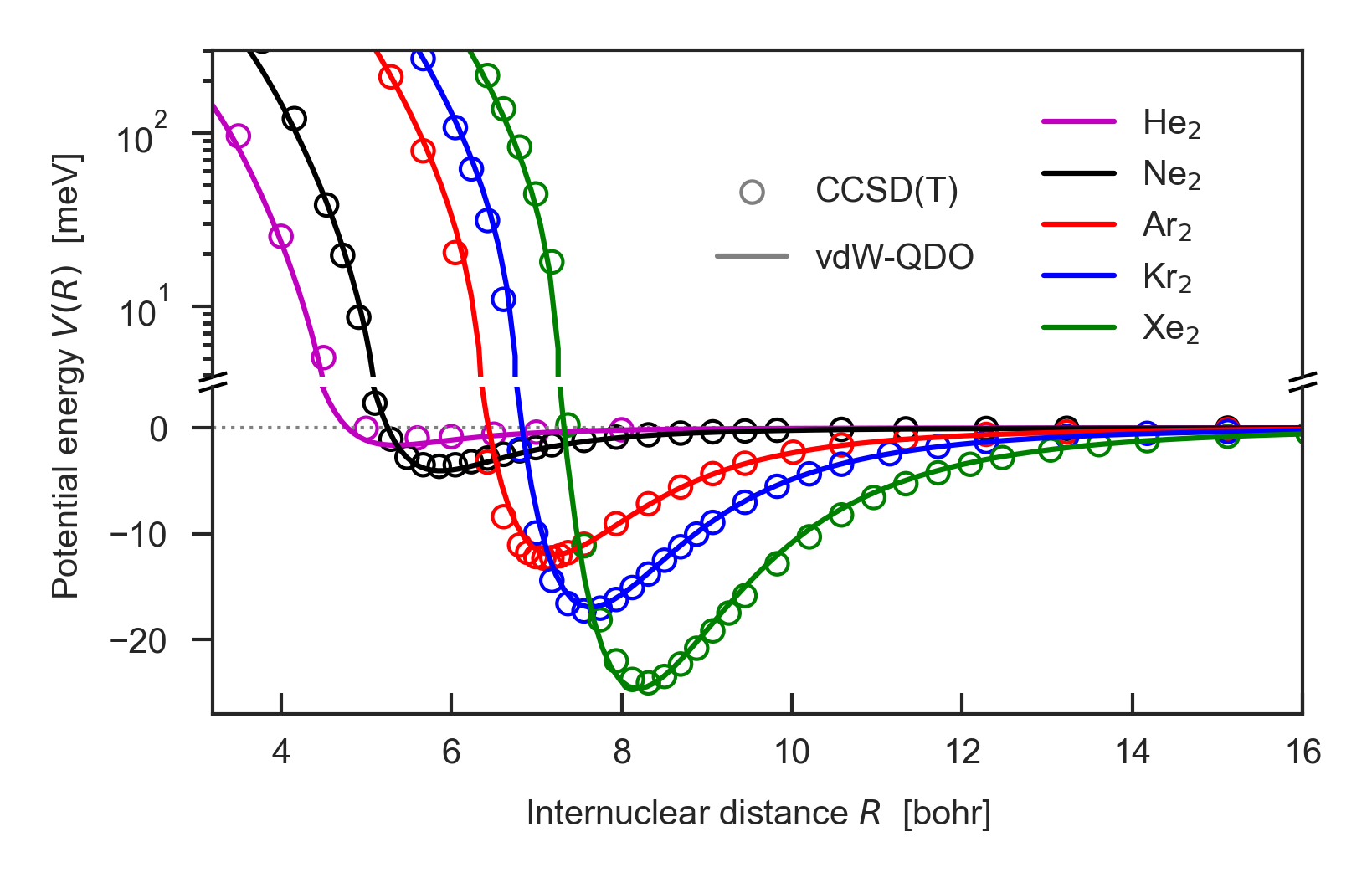}
%The surrounding frame is 9\,cm by 3.5\,cm, which is the maximum permitted for  \emph{Journal of the American Chemical Society} graphical table of content entries. The box will not resize if the content is too big: instead it will overflow the edge of the box.
%
%This box and the associated title will always be printed on a separate page at the end of the document.
%
\end{tocentry}

\newpage
\begin{abstract}
Repulsive short-range and attractive long-range van der Waals (vdW) forces have an appreciable role in the behavior of extended molecular systems. When using empirical force fields -- the most popular computational methods applied to such systems --  vdW forces are typically described by Lennard-Jones-like potentials, which unfortunately have a limited predictive power. Here, we present a universal parameterization of a quantum-mechanical vdW potential, which requires only two free-atom properties -- the static dipole polarizability $\alpha_1$ and the dipole-dipole $C_6$ dispersion coefficient. This is achieved by deriving the functional form of the potential from the quantum Drude oscillator (QDO) model, employing scaling laws for the 
equilibrium distance and the binding energy as well as applying the microscopic law of corresponding states. The vdW-QDO potential is shown to be accurate for vdW binding energy curves, as demonstrated by comparing to \emph{ab initio} binding curves of 21 noble-gas dimers. The functional form of the vdW-QDO potential has the correct asymptotic behavior both at zero and infinite distances. In addition, it is shown that the damped vdW-QDO potential can accurately describe vdW interactions in dimers consisting of group II elements. Finally, we demonstrate the applicability of the atom-in-molecule vdW-QDO model for predicting accurate dispersion energies for molecular systems. The present work makes an important step towards constructing universal vdW potentials, which could benefit (bio)molecular computational studies.
\end{abstract}

% Table of Content
%\begin{figure}[h!]
%\includegraphics[width = 3.5 in]{TOC.png}
%\caption{}
%\label{figTOC}
%\end{figure}

\newpage
\section{Introduction}

van der Waals (vdW) forces play an indisputably important role in determining the structure and dynamics of many biomolecular, solid-state and polymeric  systems~\cite{Stone2013book,Kaplan2006book,Israelachvili2011,Hermann2017-Chemrev,Stoehr2019review}. The accurate description of vdW interactions requires sophisticated quantum-mechanical treatment, using the adiabatic-connection fluctuation-dissipation theorem (ACFDT) in density-functional theory or high-level quantum chemistry methods, such as coupled cluster or quantum Monte-Carlo~\cite{Hermann2017-Chemrev,Stoehr2019review}.
However, the prohibitive computational cost of these methods precludes their applicability to extended (bio)molecular systems. Therefore, practical simulations of large and complex systems are often done using classical force fields such as AMBER~\cite{Case2005}, CHARMM~\cite{Brooks2009}, or GROMACS~\cite{Hess2008}.

For the description of vdW forces, these popular force fields resort to the seminal Lennard-Jones (LJ)~\cite{Jones1924} (or an improved Buckingham~\cite{Buckingham1938}) potential as a practical shortcut. Two parameters, the well depth $D_e$ and the equilibrium position $R_e$, fully specify the LJ potential. However, these parameters can be determined unambiguously only for relatively simple vdW-bonded systems, such as noble gas dimers or crystals. Moreover, the LJ potential is notorious for its lack of flexibility and very limited quantitative accuracy~\cite{Barker1971, Blaney1976}. On the other hand, the celebrated Tang-Toennies potentials~\cite{TangToennies2003,TangToennies2008Hg,TangToennies2020,TangToennies2021} are derived from first principles and yield high accuracy for dimers including noble gases and group II elements. To achieve such an accuracy, the Tang-Toennies potentials employ from 5 to 9 parameters, depending on the exact flavor~\cite{TangToennies2021}. 
Setting these parameters requires knowledge of $R_e$ and $D_e$ for each vdW bonded dimer~\cite{TangToennies2003,TangToennies2020}, which prevents a generalization of the Tang-Toennies models to the whole periodic table. Moreover, like the LJ potential, the most recent conformal Tang-Toennies (TTS) potential~\cite{TangToennies2020} is prone to large errors for dispersion coefficients (see Fig.~\ref{fig.: neon}a) despite its high accuracy close to equilibrium distances. Hence, a vdW potential combining wide transferability across the periodic table, high accuracy and minimal parametrization is not yet available.

Here, we develop a universal conformal pairwise vdW potential, which can be parametrized for all chemical elements based solely on two \emph{non-bonded} atomic properties -- static dipole polarizability $\alpha_1$ and dipole-dipole dispersion coefficient $C_6$. Our potential is consistently derived within the framework of the quantum Drude oscillator (QDO) model~\cite{Jones2013} using the Heitler-London perturbation theory~\cite{HeitlerLondon1927, Slater1965}, and it is devoid of adjustable parameters. This is achieved by building connections between atomic scaling laws~\cite{Fedorov2018,Vaccarelli2021,Tkatchenko2021}, the microscopic law of corresponding states~\cite{deBoer1938, deBoer1948, Mcquarrie1997,Rowlinson2002}, and symmetry-adapted perturbation theory (SAPT)~\cite{Szalewicz1994-ChemRev,Szalewicz2022} for intermolecular interactions. The derived exchange repulsion term in our potential obeys correct physical limits both at $R \rightarrow 0$ and $R \rightarrow \infty$, and the predicted $C_6$ dispersion coefficients are significantly more accurate compared to the other conformal Lennard-Jones and Tang-Toennies~\cite{TangToennies2020} potentials. The designed vdW-QDO potentials are twice as accurate as the LJ potentials, when averaged over 15 noble-gas dimers. In addition, the vdW-QDO potential augmented by a damping function can accurately describe binding curves of dimers consisting of (closed-shell) group II atoms. Moreover, the vdW-QDO potential can be applied to molecular systems, when coupled with atom-in-molecule (AIM) approach~\cite{Tkatchenko2009}. We demonstrate this by accurately reproducing the dispersion energy for dispersion-dominated molecular dimers from the S66$\times$8 dataset~\cite{Rezac2011}.

We derive the vdW potential in the QDO framework, which is a coarse-grained model for the electronic response~\cite{Bade1957, Thole1981, Wang2001, Sommerfeld2005, Jones2010thesis, Martyna2019-RevModPhys} proved to be accurate and insightful in many applications across various fields~\cite{Whitfield2006, Martyna2013-water, Sokhan2015-water, Tkatchenko2012, Jones2013, Ambrosetti2014, Martyna2019-RevModPhys, Vaccarelli2021, Ambrosetti2022, Karimpour2022, Karimpour2022_JPCL, Tkatchenko2023}. Within the QDO model, the response of valence electrons is described \emph{via} a quasi-particle (\emph{drudon} or Drude particle) with a negative charge $-q$ and mass $\mu$, harmonically bound to a positively-charged pseudo-nucleus of charge $q$ with a characteristic frequency $\omega$. Coupled QDOs are also extensively used in the development of vdW density functionals~\cite{Tkatchenko2009, 
Tkatchenko2012, Ambrosetti2012-QDO}, quantum mechanical~\cite{Jones2013, Martyna2019-RevModPhys} and polarizable force fields~\cite{Harder2006, Lopes2013, Adluri2015, Sokhan2015-water, Piquemal2022-MBD1} as well as recent machine learning force fields~\cite{Muhli2021, Piquemal2022-MBD2}. 
 
The QDO model has been already used to build interatomic vdW potentials for water or noble-gas dimers and crystals~\cite{Jones2013,Martyna2013-water,Sokhan2015-water,Sadhukhan2016,Martyna2019-RevModPhys}. However, within the corresponding studies, the repulsive term was added in \emph{ad hoc} manner, either by fitting Born-Mayer~\cite{BornMayer1932, Buckingham1938} exponents to \emph{ab initio} repulsive walls~\cite{Jones2013,Martyna2013-water,Sokhan2015-water,Martyna2019-RevModPhys} or by directly adding the Hartree-Fock exchange energy~\cite{Sadhukhan2016}. Therefore, such potentials cannot be generalized beyond the systems for which direct first-principles simulations are possible. In contrast, here we suggest a consistent treatment of both Pauli (exchange) repulsion and vdW dispersion within the QDO framework. To our knowledge, this is the first vdW potential of such a type, which does not directly utilize the reference binding energy of dimers or the Hartree-Fock exchange energy curve, but nevertheless provides relatively good accuracy.

\section{Results}
\subsection{Model construction}

The long-range vdW dispersion energy for two identical QDOs is given by the usual multipolar series~\cite{Jones2010thesis,Jones2013}
\begin{equation}
 E_{\rm disp}(R) = -\frac{C_6}{R^6} -\frac{C_8}{R^8} -\frac{C_{10}}{R^{10}} - ... \ ,
\label{eq.: Edisp}
\end{equation}
where the dispersion coefficients are related to the oscillator parameters \emph{via} the closed-form expressions~\cite{Jones2013}
\begin{equation}
 C_6 = \frac{3}{4} \hbar \omega \alpha_1^2 k_e^2\ , \ C_8 = \frac{5 \hbar}{\mu \omega}\, C_6\ , \ C_{10} = \frac{245 \hbar^2}{8 (\mu \omega)^2}\, C_6\ ,
\label{eq.: C2n_qdo}
\end{equation}
where $\alpha_1 = q^2/\mu\omega^2$ is the QDO dipole polarizability and $k_e = 1/4\pi\varepsilon_0$. Tang and Toennies showed~\cite{TangToennies2003, TangToennies2008Hg} that including the three leading dispersion terms is sufficient to obtain the accurate vdW potential. Therefore, we also truncate the series of Eq.~(\ref{eq.: Edisp}) at the $C_{10}$ term.

\begin{figure*}[t!]
\includegraphics[width=1.2\LL]{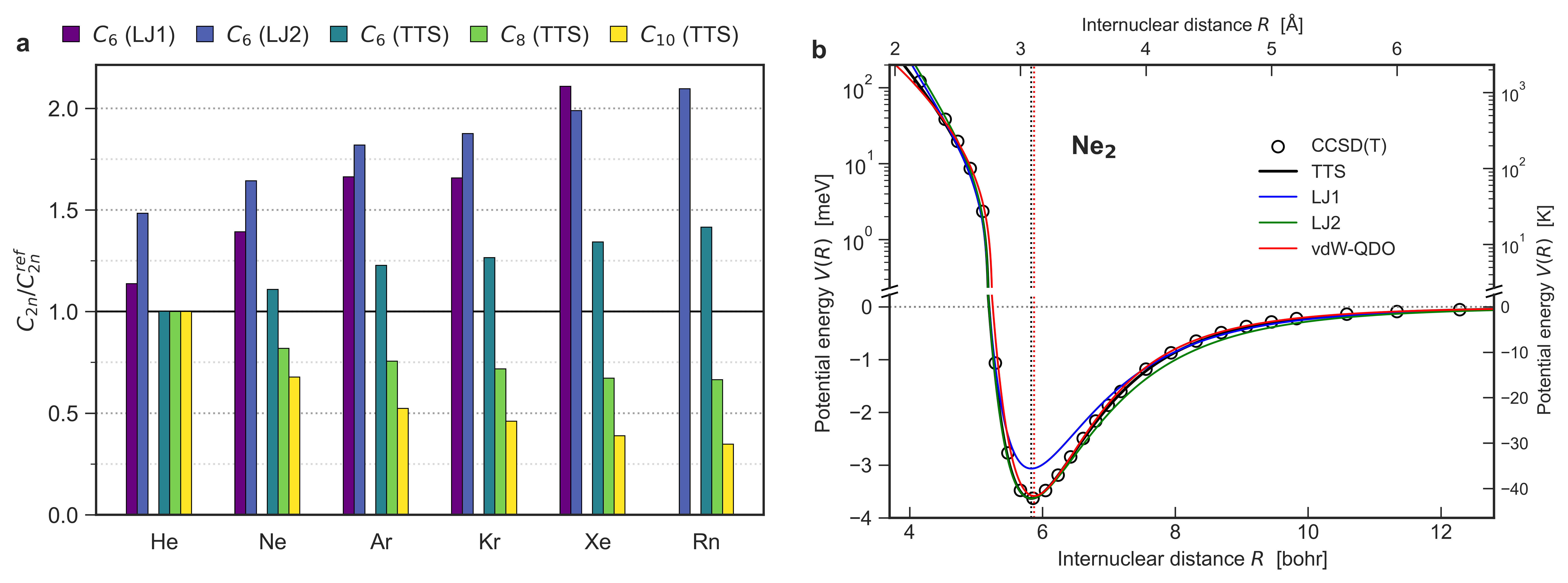}
\caption{\footnotesize (a) Errors in dispersion coefficients arising from the LJ and TTS potentials. The TTS dispersion coefficients are obtained as $C_{2n} = C_{2n}^* \times D_eR_e^{2n}$~\cite{TangToennies2020}, and the reference dispersion coefficients $C_{2n}^{\rm ref}$ are given in Table~\ref{tab:ngrefvalues}. (b) The vdW-QDO potential for neon dimer benchmarked to the TTS potential and the reference CCSD(T) potential~\cite{Hellmann2008}. Vertical dotted lines indicate the equilibrium distance as predicted by CCSD(T) (black) and vdW-QDO potential (red). For the comparison, the LJ potentials in two different parametrizations (see our discussion in the text) are also displayed in blue and green.}
\label{fig.: neon}
\end{figure*}

The exchange repulsion is introduced into the model according to Refs.~\citenum{Fedorov2018, Vaccarelli2021}, where multipole contributions to the exchange energy of a homonuclear dimer were derived by considering two identical drudons as bosons, assuming that they represent closed valence shells of atoms with zero total spin. Consequently, the total wave function of a dimer is represented by a permanent
\begin{equation}
\Psi({\mathbf r_1}, {\mathbf r_2}) = \tfrac{1}{\sqrt{2}}\left(\psi_A(\mathbf r_1)\psi_B(\mathbf r_2) + \psi_A(\mathbf r_2)\psi_B(\mathbf r_1)\right)\ ,
\label{eq.: WFperm}
\end{equation}
where $\psi_A(\mathbf r_1) = (\mu\omega/\pi\hbar)^{\nicefrac{3}{4}}\, e^{-(\mu\omega/2\hbar) {\mathbf r^2_1}}$ and $\psi_B(\mathbf r_2) = (\mu\omega/\pi\hbar)^{\nicefrac{3}{4}}\, e^{-(\mu\omega/2\hbar) (\mathbf r_2 - \mathbf R)^2}$ are,~respectively, the ground-state wave functions of drudons centered at nuclei $A$ and $B$ separated by $\mathbf{R}$.  Within the Heitler-London perturbation theory~\cite{HeitlerLondon1927,Slater1965}, the exchange energy of two identical vdW-bonded QDOs for distances near equilibrium and larger is well approximated by the exchange integral~\cite{Fedorov2018}
\begin{equation}
 E_{\rm ex} \approx J_{\rm ex} = \bra{\psi_A(\mathbf r_1)\psi_B(\mathbf r_2)} \hat{V}_C \ket{\psi_A(\mathbf r_2)\psi_B(\mathbf r_1)}\ .
\label{eq.: Jex}
\end{equation}
The evaluation of Eq.~(\ref{eq.: Jex}) with the multipole expansion of Coulomb coupling $\hat{V}_C$ between the two QDOs results in multipole contributions to the exchange energy~\cite{Fedorov2018, Vaccarelli2021}. 
In dipole approximation, this yields
\begin{equation}
J_{\rm ex}^{(1)} = k_e q^2 S/2R\ ,\ \ 
S = | \braket{\psi_A}{\psi_B} |^2 = e^{-\frac{\mu\omega}{2\hbar}R^2}\ ,
\label{eq.: Jexdip}
\end{equation}
where $S$ is the overlap integral. Higher-order multipole contributions ($l > 1$) to the exchange repulsion energy $J_{\rm ex}^{(l)}$ have the same leading-term dependence on internuclear distance $R$, with the only difference in a proportionality coefficient, i.e. $J_{\rm ex}^{(l)} \propto k_e q^2 S/R$~\cite{Vaccarelli2021}. Therefore, we introduce an effective exchange repulsion energy as 
\begin{equation}
E^{\rm eff}_{\rm ex} = A k_e q^2 S/R\ ,
\label{eq.: JexTot}
\end{equation}
with the proportionality coefficient $A$ to be determined self-consistently, in what follows. In this way, we effectively include multipole contributions to all orders. Importantly, our $E^{\rm eff}_{\rm ex}$ has $1/R$ dependence, which properly describes the infinite repulsive wall at short distances. Thus, in contrast to the Born-Mayer or Duman-Smirnov~\cite{DumanSmirnov1970, Kleinekathofer1995,TangToennies1998} functional forms for exchange repulsion possessing a finite value of $E_{\rm ex}$ at $R \rightarrow 0$, Eq.~\eqref{eq.: JexTot} is in agreement with the orbital overlap model for Pauli repulsion~\cite{Salem1961,Rackers2019}. Moreover, our $E^{\rm eff}_{\rm ex}$ does not rely on  empiricism, as it explicitly depends only on the QDO parameters (\emph{vide infra}), whereas the existing Pauli repulsion models require fitting to some \emph{ab initio} data~\cite{TangToennies1986,TangToennies1998,Whitfield2006,Jones2013,Vanvleet2016,Martyna2019-RevModPhys,Rackers2019}.

In order to determine the coefficient $A$ in Eq.~\eqref{eq.: JexTot}, we employ the force balance condition at the equilibrium distance, $\left( -\nabla_R  E_{\rm ex}^{\rm eff} - \nabla_R E_{\rm disp} \right) |_{R = R_e} = 0$, which yields
\begin{align}
 Ak_e q^2 \left[ \frac{1}{R_e^2} + \frac{\mu\omega}{\hbar} \right] e^{-\frac{\mu\omega}{2\hbar}R_e^2} = \frac{6C_6}{R_e^7} + \frac{8C_8}{R_e^9} + \frac{10C_{10}}{R_e^{11}}\ .
\label{eq.: forcebalance}
\end{align}
To evaluate the equilibrium distance $R_e$ in our model, we use the quantum-mechanical relation between the atomic (static) dipole polarizability and vdW radius~\cite{Fedorov2018}
\begin{equation}
\alpha_1 = \Phi \times R_{\rm vdW}^7\ ,
\label{eq.: Rvdw}
\end{equation}
where the proportionality coefficient $\Phi$ is given by~\cite{Tkatchenko2021}
\begin{equation}
\Phi = (4\pi\varepsilon_0/a_0^4) \times \alpha_{\rm fsc}^{4/3}\ ,
\label{eq.: Phiconst}
\end{equation}
with $\alpha_{\rm fsc} = e^2/4\pi\varepsilon_0\hbar c \approx 1/137.036$ as the fine-structure constant. The relation given by Eqs.~(\ref{eq.: Rvdw})--(\ref{eq.: Phiconst}) turned out to be valid for real atoms. Especially, it is very accurate for noble gases, where the mean absolute relative error (MARE) $\langle | R_{\rm vdW} - R_{\rm vdW}^{\rm ref} | / R_{\rm vdW}^{\rm ref} \rangle$ is about 1\%\,.~\cite{Fedorov2018, Tkatchenko2021} Since by definition $R_{\rm vdW}$ is a half of the equilibrium distance $R_e$ in a homonuclear vdW bonded dimer~\cite{Bondi1964, Fedorov2018}, accurate equilibrium distances can be obtained \emph{via}
\begin{equation}
R_e = 2 \times R_{\rm vdW} = 2 \times (\alpha_1/\Phi)^{1/7}\ .
\label{eq.: eqdist}
\end{equation}
With $\alpha_1$ and $C_6$ being fixed, there are two unknown quantities in Eq.~(\ref{eq.: forcebalance}), $A$ and $\mu\omega$, since $C_8$ and $C_{10}$ are solely expressed in terms of $C_6$ and $\mu\omega$ via Eq.~(\ref{eq.: C2n_qdo}).

As shown in Ref.~\citenum{Tkatchenko2023}, the product $\mu\omega$ can be obtained from the force balance in the dipole approximation
\begin{equation}
 \frac{k_e q^2}{2} \left[ \frac{1}{R_e^2} + \frac{\mu\omega}{\hbar} \right] e^{-\frac{\mu\omega}{2\hbar}R_e^2} = \frac{6C_6}{R_e^7}\ ,
\label{eq.: dipforcebalance}
\end{equation} 
with $R_e$ substituted from Eq.~(\ref{eq.: eqdist}). Solution of this
transcendental equation allows to determine the three oscillator parameters $\{ q, \mu, \omega \}$ given only $\{ \alpha_1, C_6 \}$. Let us call this parametrization scheme vdW-OQDO, similar to the recently suggested OQDO scheme~\cite{Tkatchenko2023}. The details of the procedure and the corresponding values of $\{ q, \mu, \omega \}$ can be found in the Supporting Information. 

Solving Eqs.~(\ref{eq.: dipforcebalance}) and (\ref{eq.: forcebalance}) together, one can obtain
\begin{equation}
 A = \frac{1}{2} + \frac{2C_8}{3C_6 R_e^2} + \frac{5C_{10}}{6C_6 R_e^4}
\label{eq.: Acoeff}
\end{equation}
and the total vdW potential
\begin{equation}
 V_{\rm QDO} = A \frac{k_e q^2}{R} e^{-\frac{(\gamma R)^2}{2}} - \sum_{n = 3}^{5} \frac{C_{2n}}{R^{2n}}\ , \ \gamma = \sqrt{\mu\omega/\hbar} \ .
\label{eq.: Vvdwtot}
\end{equation}

\begin{table*}[t!]
\footnotesize
\caption{\footnotesize The reference values of static dipole polarizability $\alpha_1$ (in a.u.), dispersion coefficients $C_6, C_8, C_{10}$ (in a.u.), and dimer potential well parameters $R_e$ (in bohr) and $D_e$ (in meV) for noble-gas dimers. For $R_e$ and $D_e$, the values in \AA~and Kelvin, respectively, are extra given in parentheses. Also, we compare their reference values (columns 5 and 7) to the predictions of Eqs.~(\ref{eq.: eqdist}) and (\ref{eq.: Descale}) (columns 6 and 8). The values for $C_6$, $C_8$, and $C_{10}$ labeled with the star ($*$) are taken from Ref.~\citenum{TangToennies2003} instead of Refs.~\citenum{Gobre2016,Jiang2015}.}
\label{tab:ngrefvalues}
\begin{tabular}{|c|c c c c c c c c|}
\hline
~~ & $\alpha_1$~\cite{Gobre2016} & $C_6$~\cite{Gobre2016} & $C_8$~\cite{Jiang2015} & $C_{10}$~\cite{Jiang2015} & $R_e^{\rm ref}$ (in \AA)~\cite{TangToennies2020} & $R_e$ (in \AA) & $D_e^{\rm ref}$ (in K)~\cite{TangToennies2020} & $D_e$ (in K) \\
\hline
 He$_2$ & 1.38 & 1.46 & 14.123 & 183.79 & 5.608~(2.97) & 5.35~(2.83) & 0.948~(10.99) & 1.634~(19.0) \\
 Ne$_2$ & 2.67 & 6.38 & 90.265 & 1532.8 & 5.83~~(3.09) & 5.87~(3.11) & 3.632~(42.15) & 4.049~(47.0) \\
 Ar$_2$ & 11.1 & 64.3 & 1621.5 & 49033 & 7.11~~(3.76) & 7.20~(3.81) & 12.319~(142.95) & 12.00~(139.3) \\
 Kr$_2$ & 16.8 & 129.6 & 4040 & 150130 & 7.589~(4.02) & 7.64~(4.04) & 17.310~(200.87) & 16.94~(196.6) \\
 Xe$_2$ & 27.3 & 285.9 & 12004 & 588210 & 8.273~(4.38) & 8.19~(4.33) & 24.126~(279.97) & 24.64~(285.9) \\
 Rn$_2$ & 33.54 & 420.6$^{*}$ & 19263$^{*}$ & 1067000$^{*}$ & 8.37~~(4.43) & 8.43~(4.46) & 34.885~(404.81) & 30.38~(352.5) \\
\hline
\end{tabular}
\end{table*}

The vdW-QDO potential for neon is displayed by the red curve in Fig.~\ref{fig.: neon}b, which shows excellent agreement with the TTS potential~\cite{TangToennies2020} as well as with the CCSD(T) calculations~\cite{Hellmann2008} across the whole range of distances from $0.7R_e$ ($\sim$4 bohr) to infinity. Inclusion of $C_8$ and $C_{10}$ dispersion coefficients together with the suggested approach to treat exchange repulsion energy allows us to predict the correct depth and shape of the potential without losing the accuracy in predicting the equilibrium distance, which is inherited from the dipole approximation. In addition, we compare our potential to the LJ potential, for which we use two different parametrizations: LJ1 derived from thermodynamical properties and LJ2 designed to reproduce reference $R_e$ and $D_e$ (see the Supporting Information for more details). We note that the present vdW-QDO potential~(\ref{eq.: Vvdwtot}) performs accurately in the whole range of distances, whereas the LJ1 potential (blue curve in Fig.~\ref{fig.: neon}b) underestimates the energy in potential minimum region and the LJ2 potential overestimates the long-range energy (green curve), although both being reasonably accurate in the repulsive region. This imbalance and lack of flexibility of the LJ potential, which is observed for all noble gases, is one of the main issues limiting its quantitative predictive power~\cite{Barker1971, Blaney1976}. Moreover, the LJ potential severely overestimates $C_6$ coefficient~(Fig.~\ref{fig.: neon}a), which is responsible for the correct long-range energy. The proposed vdW-QDO potential overcomes these difficulties without increasing the number of parameters. Moreover, our potential recovers correct bonding behavior using only a free atom property $\alpha_1$ and asymptotic interaction parameter $C_6$, which do not contain information about the interaction between atoms at short distances.

\subsection{Application to noble-gas dimers}

With the accurate Ne$_2$ potential curve in hand, its counterparts for all other noble-gas dimers can be derived using the conformality of their potentials~\cite{TangToennies2020,Farrar1973,TangToennies1977}, which is a microscopic manifestation of the law of corresponding states~\cite{deBoer1938,deBoer1948,Mcquarrie1997}. Namely, for the vdW potential of other noble-gas dimers we write
\begin{equation}
 V_{\rm QDO}(R) = D_e U_{\rm QDO}^{\rm Ne}(x), \ x = {R}/{R_e}\ ,
\label{eq.: Vscaled}
\end{equation}
where $U_{\rm QDO}^{\rm Ne}(x) = V_{\rm QDO}^{\rm Ne}(xR_e^{\rm Ne})/V_{\rm QDO}^{\rm Ne}(R_e^{\rm Ne})$ is the dimensionless potential (shape) of Ne$_2$ dimer
\begin{equation}
\label{eq.: U_Ne2}
   U_{\rm QDO}^{\rm Ne}(x) =  \frac{A^*}{x} e^{-\frac{(\gamma^* x)^2}{2}} - \sum_{n=3}^{5} \frac{C_{2n}^*}{x^{2n}}\ ,
\end{equation}
with the numerical values of the starred (unitless) parameters and their definitions presented in Table~\ref{tab:ngpotpar}.

\begin{table}[!h]
\footnotesize
\caption{\footnotesize The dimensionless parameters in Eq.~(\ref{eq.: U_Ne2}). The neon dimer parameters used in the second column are $D_e({\rm Ne}_2)~=~3.586 \ {\rm meV} = 13.178 \times 10^{-5}~{\rm a.u.}$ [Eq.~(\ref{eq.: Defull})] and $R_e({\rm Ne}_2) = 5.875$ bohr [Eq.~(\ref{eq.: eqdist})]. The QDO parameters for Ne$_2$ dimer are $q~=~1.18865$, $\mu = 0.37164$, $\omega = 1.19326$ (all in a.u.).}
\label{tab:ngpotpar}
\begin{center}
\begin{tabular}{|c|c|c|}
\hline
~~Parameter~~ & ~Definition~ & ~~Numerical value~~ \\
\hline
~$A^*$~ & ~$A k_e q^2/R_e D_e$~ & ~1508.917~ \\
~$\gamma^*$~ & ~$R_e\sqrt{\mu\omega/\hbar}$~ & ~3.912~ \\
~$C_6^*$~ & ~$C_6/D_eR_e^6$~ & ~1.1779~ \\
~$C_8^*$~ & ~$5C_6/D_eR_e^6(\gamma^*)^2$~ & ~0.3848~ \\
~$C_{10}^*$~ & ~~$245C_6/8D_eR_e^6(\gamma^*)^4$~~ & ~0.1540~ \\
\hline
\end{tabular}
\end{center}
\end{table}

Thus, only $R_e$ and $D_e$ for every dimer are required to obtain their vdW potential. For $R_e$, the accurate scaling law~(\ref{eq.: eqdist}) is already established, whereas an analogous scaling law for $D_e$ of noble-gas dimers is not yet known. Substituting $R = R_e$ to Eq.~(\ref{eq.: Vvdwtot}) and using Eq.~(\ref{eq.: forcebalance}) to eliminate $A k_e q^2 e^{-\frac{(\gamma R)^2}{2}}/R$ yields
\begin{equation}
\label{eq.: Defull}
 D_e = -V_{\rm QDO}^{\rm Ne}(R_e^{\rm Ne}) = \frac{C_6}{R_e^6} \left( 1 - \frac{\beta - 5}{\beta (1 + \beta)} -\frac{40}{\beta (1 + \beta)} 
  + \frac{245}{8\beta^2} - \frac{2450}{8\beta^2(1 + \beta)} \right) \ , 
\end{equation}
with $\beta = \frac{\mu\omega}{\hbar} R_e^2 = (\gamma^*)^2$. Analyzing reference CCSD(T) data for $D_e$ from Ref.~\citenum{TangToennies2020}, we found that Eq.~(\ref{eq.: Defull}) truncated at first two terms can accurately predict $D_e$ for all noble-gas dimers
\begin{equation}
 D_e \approx \frac{C_6}{R_e^6} \left( 1 - \frac{\beta - 5}{\beta (1 + \beta)} \right)\ .
\label{eq.: Descale}
\end{equation}

\begin{figure}[h]
\includegraphics[width=0.7\LL]{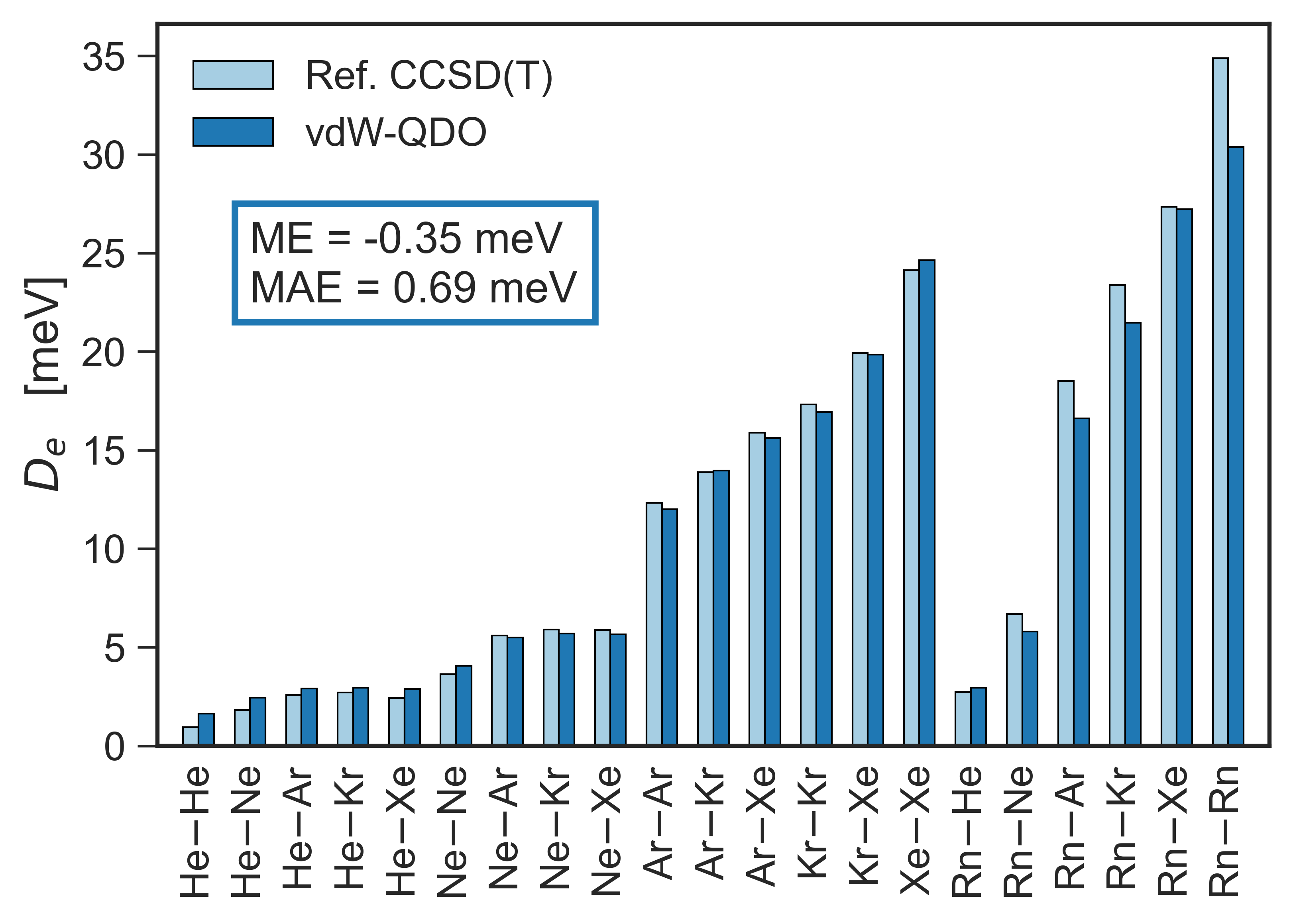}
\caption{\footnotesize Potential well depth $D_e$ by Eq.~(\ref{eq.: Descale}) compared to the reference CCSD(T) values~\cite{TangToennies2020} for 21 noble-gas dimers. Mean error (ME) and mean absolute error (MAE) are displayed.}
\label{fig.: De}
\end{figure}

\begin{figure*}[t!]
\includegraphics[width=1.2\LL]{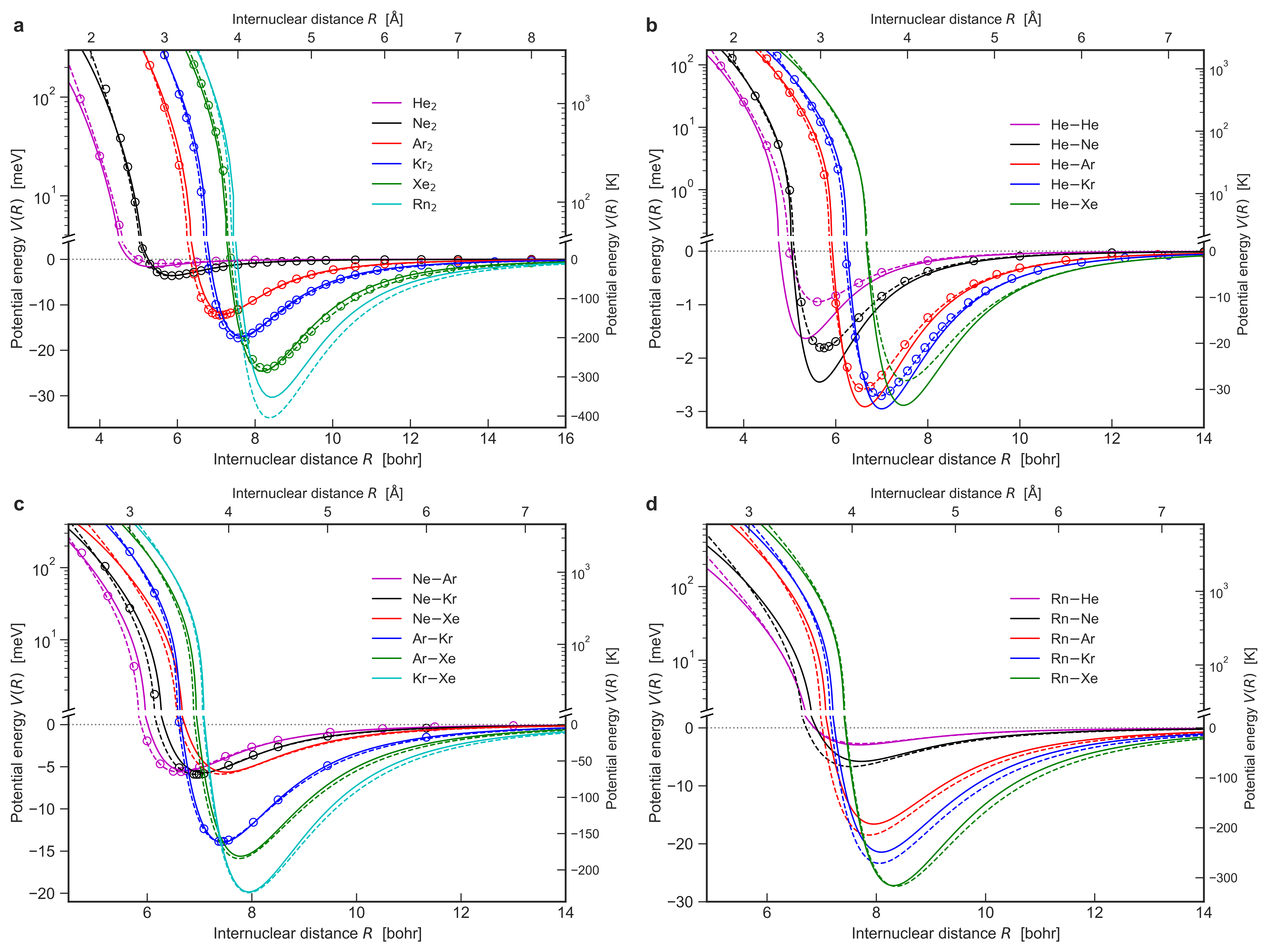}
\caption{\footnotesize vdW-QDO potentials (solid lines) for (a) homonuclear and (b-d) heteronuclear noble-gas dimers benchmarked to the TTS potential~\cite{TangToennies2020} (dashed lines) and the reference CCSD(T) calculations (circles)~\cite{Cencek2012, Hellmann2008, Patkowski2010, Jager2016, Hellmann2017, Lopez2004, Haley2003}.}
\label{fig.: dimers}
\end{figure*}

In Fig.~\ref{fig.: De}, $D_e$ by Eq.~(\ref{eq.: Descale}) are compared to the reference CCSD(T) data. The bar chart shows that Eq.~(\ref{eq.: Descale}) is accurate for homo- and heteronuclear dimers of He--Xe with all errors below 1 meV. For dimers with Rn the error is larger, with Rn$_2$ being underbound by 4.5 meV, or 13\%. The larger errors for Rn dimers likely stem from the fact that the reference coupled-cluster calculation~\cite{Shee2015} is less reliable than the corresponding calculations for the lighter dimers He$_2$ -- Xe$_2$~\cite{Cencek2012,Hellmann2008,Patkowski2010,Jager2016,Hellmann2017}. For example, $D_e$ of Xe$_2$ dimer reported in Ref.~\citenum{Shee2015} is by 7.5\% larger than the one of Ref.~\citenum{Hellmann2017}, which is state-of-the-art calculation. Thus, similar or even larger overestimation of $D_e$ should be expected for Rn$_2$~\cite{Shee2015}, where relativistic effects are more pronounced. Accounting for that, the estimated error of Eq.~(\ref{eq.: Descale}) for Rn$_2$ would not exceed 5.5\%. We conclude that the suggested scaling law~(\ref{eq.: Descale}) allows one to accurately evaluate $D_e$ for all noble-gas dimers given only $\{ \alpha_1, C_6 \}$ without involving any adjustable parameters.

To extend the developed potential to heteronuclear dimers, combination rules for potential parameters can be used. The simplest ones are given by
\begin{equation}
 R_e^{AB} = \left(R_e^{A} + R_e^{B}\right)/2\ , \ \ D_e^{AB} = \sqrt{D_e^{A} D_e^{B}}
\label{eq.: LorentzBerthelot}
\end{equation}
and known as the Lorentz-Berthelot rules, which are often used for the LJ potential and implemented in many molecular simulation packages~\cite{Case2005, Brooks2009, Hess2008}. However, the Lorentz-Berthelot rules are not accurate~\cite{Delhommelle2001,Song2003,Boda2008,Fedorov2018}. 
Therefore, instead of mixing $R_e$ and $D_e$, we use mixing rules for $\alpha_1$ and $C_6$, since our potential is fully parametrized by these two quantities. With the effective mixed values $\{ \alpha_1^{AB}, C_6^{AB} \}$, we can set three oscillator parameters $\{ q,\mu, \omega \}$ through the same vdW-OQDO parametrization procedure as for homonuclear dimers, which is described in the Supporting Information. By doing so, even the heteronuclear dimer AB is effectively represented by two \emph{identical} oscillators, which still allows us to apply the formalism for exchange repulsion, Eqs.~(\ref{eq.: WFperm})--(\ref{eq.: Jex}), developed for homonuclear dimers. For $C_6$ dispersion coefficient, the very accurate combination rule arising from the London formula is already well known~\cite{Tang1969, TangToennies2003}
\begin{equation}
 C_6^{AB} = \frac{2\alpha_1^{A} \alpha_1^{B} C_6^{A} C_6^{B}}{C_6^{A} (\alpha_1^{B})^2 + C_6^{B}(\alpha_1^{A})^2}\ .
\label{eq.: c6mix}
\end{equation}

To combine polarizabilities, we employ the robust mixing rule for vdW radii which was established in Ref.~\citenum{Fedorov2018}, where it was shown that accurate equilibrium distances in noble-gas dimers (MARE = 1\%) are 
delivered by
\begin{equation}
 R_e^{AB} = 2 \times \Phi^{-1/7} \left[ \left(\alpha_1^A + \alpha_1^B\right)/2 \right]^{1/7}\ ,
\label{eq.: Remix}
\end{equation} 
similar to the homonuclear case~(\ref{eq.: Rvdw}). Thus, the effective polarizability $\alpha_1^{AB}$ can be simply represented by
\begin{equation}
 \alpha_1^{AB} = \left(\alpha_1^A + \alpha_1^B\right)/2\ .
\label{eq.: alphamix}
\end{equation}
Thereby, combining Eqs.~(\ref{eq.: Vscaled}), (\ref{eq.: U_Ne2}), (\ref{eq.: Descale}), (\ref{eq.: c6mix}) and~(\ref{eq.: alphamix}) with the vdW-OQDO parametrization scheme (see the Supporting Information), we obtain vdW-QDO potentials for all 21 noble-gas dimers. They are shown in Fig.~\ref{fig.: dimers} with excellent agreement to both TTS potential and reference CCSD(T) calculations for homo- and heteronuclear dimers of Ne, Ar, Kr and Xe on panels (a) and (c), as well as for He--Ar, He--Kr (b) and Rn--Xe (d). The other Rn dimers are challenging for our model due to the discrepancies in $D_e$, as was discussed above. In case of He dimers, to large extent the error is caused by the underestimated $R_e$ of He$_2$, with 5.35 bohr predicted by Eq.~(\ref{eq.: eqdist}) against the reference value of 5.608 bohr~\cite{Cencek2012}. In addition, the actual error in potential for He dimers is small in magnitude (despite being seemingly large visually due to the scale of $y$-axis in Fig.~\ref{fig.: dimers}b). 

\begin{figure*}[t!]
\includegraphics[width=1.2\LL]{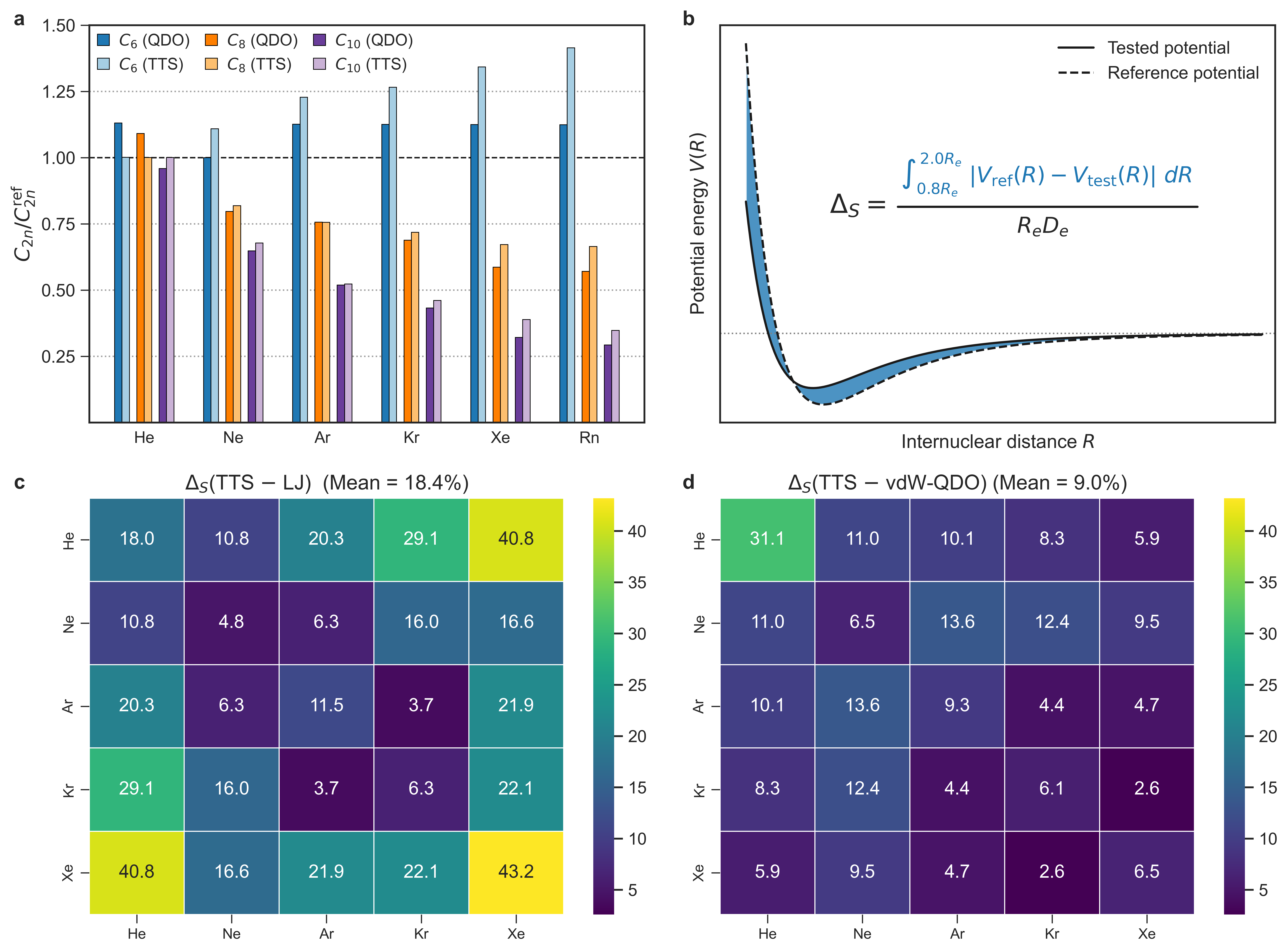}
\caption{\footnotesize (a) Errors in dispersion coefficients predicted by vdW-QDO (dark colors) and TTS (light colors) potentials. (b) Schematic illustration of the $\Delta_S$ metric calculation. (c) Heatmaps showing $\Delta_S$ (in \%) for LJ1 (left) and vdW-QDO (right) potentials with respect to the reference TTS potential. The left and right colorbars have the same scale.}
\label{fig.: errors}
\end{figure*}

To evaluate the accuracy of our potential quantitatively, we introduce the normalized area difference metric $\Delta_S$ between tested and reference potentials as
\begin{equation}
 \Delta_S = \frac{1}{R_e D_e} \int_{0.8R_e}^{2.0R_e}~\left| V_{\rm ref}(R) - V_{\rm test}(R) \right|~dR\ .
\label{eq.: deltaS}
\end{equation}
The essential physical meaning of $\Delta_S$ is illustrated in Fig.~\ref{fig.: errors}b which shows that this single unitless number represents a measure of difference between tested and reference potentials. 
The integration limits are set to $0.8R_e$ and $2.0R_e$ in order to
evaluate accuracy close to the minimum region, whereas the long-range accuracy can be evaluated separately in terms of dispersion coefficients. Previously, $\Delta$-gauge was used in benchmarks of various density-functional codes in calculations of equations of state for solids~\cite{Lejaeghere2014, Lejaeghere2016}. The TTS potential~\cite{TangToennies2020} was chosen as the reference potential, and we benchmark vdW-QDO and LJ1 potentials with respect to it (a similar benchmark for LJ2 can be found in the Supporting Information). The computed $\Delta_S$-matrices for 15 He -- Xe dimers are displayed in 
Fig.~\ref{fig.: errors}c-d (Rn dimers are omitted since there are
no LJ parameters for Rn available in literature). We note that the vdW-QDO potential has twice better accuracy than the LJ one with $\langle \Delta_S^{\rm QDO} \rangle = 9.0\%$ compared to $\langle \Delta_S^{\rm LJ1} \rangle = 18.4\%$ and $\langle \Delta_S^{\rm LJ2} \rangle = 17.3\%$ when averaged over 15 dimers. Helium is the obvious outlier for vdW-QDO with the highest $\Delta_S = 31.1\%$, whereas for all other dimers $\Delta_S$ is below 13.6\%. In contrast, LJ potential shows much broader variations in $\Delta_S$ spanning from 3.7\% for Ar--Kr to 43.2\% for Xe$_2$. Although the LJ potential shows better $\Delta_S$ values than vdW-QDO for some of the dimers (He$_2$, Ne$_2$, He--Ne, Ne--Ar, Ar--Kr), overall the performance of vdW-QDO potential is more accurate and robust. Generally, Fig.~\ref{fig.: errors} supports the above conclusions about the accuracy of vdW-QDO potential based on Fig.~\ref{fig.: dimers}. We note that the predictions of LJ potential become worse for heteronuclear dimers composed of small and large atoms 
(e.g.~He--Ar, He--Kr, Ne--Kr) than for atoms with a relatively close size (Ne--Ar, Ar--Kr) (Fig.~\ref{fig.: errors}c). In contrast, the evenly accurate predictions of the vdW-QDO model (Fig.~\ref{fig.: errors}d) suggest that the combination rules~(\ref{eq.: c6mix}) and~(\ref{eq.: alphamix}) employed in this work are more accurate and robust than the Lorentz-Berthelot rules~(\ref{eq.: LorentzBerthelot}).

To evaluate the quality of the potentials in the long-range limit, in Fig.~\ref{fig.: errors}a we compare the dispersion coefficients predicted by vdW-QDO and TTS potentials to the reference \emph{ab initio} values (Table~\ref{tab:ngrefvalues}). Such a comparison is fair, since both potentials are built as conformal ones, unlike the earlier TT-2003 potentials~\cite{TangToennies2003}, which directly utilize reference $C_{2n}$ coefficients for every element being therefore not strictly conformal.
To recover the TTS dispersion coefficients, we used the reported scaling law $C_{2n} = C_{2n}^* \times D_eR_e^{2n}$ with $C_6^* = 1.3499, \ C_8^*~=~0.4147, \ C_{10}^* = 0.1716$~\cite{TangToennies2020}. For vdW-QDO dispersion coefficients, from Eqs.~(\ref{eq.: Vscaled}), (\ref{eq.: U_Ne2}) and (\ref{eq.: Descale}) one can obtain
\begin{equation}
 C_{2n} = C_{2n}^* \times \frac{C_6}{R_e^{2n-6}} \left[ 1 - \frac{\beta - 5}{\beta(\beta+1)} \right]\ ,
\label{eq.: c2n_scale}
\end{equation}
where $C_6^* = 1.1779, \ C_8^*~=~0.3848, \ C_{10}^* = 0.1540$ (see Table~\ref{tab:ngpotpar}). We found that both potentials severely underestimate $C_8$ and $C_{10}$ and demonstrate similar magnitude of these errors increasing with the atomic number. However, for $C_6$ vdW-QDO potential performs much better, showing homogeneous overestimate of 12-13\%, whereas the TTS potential again possesses increasing magnitude of error, reaching its maximum of 41\% for Rn. While $C_8$ and $C_{10}$ are important to deliver accurate potential near the equilibrium, in the asymptotic limit the quality of the potential is fully determined by the leading $C_6$ coefficient. Therefore, we can conclude that our conformal vdW-QDO potential shows physically more reasonable long-range behavior than the conformal TTS potential.

In contrast to the Tang-Toennies potentials~\cite{TangToennies2003,TangToennies2020}, the above vdW-QDO model does not employ any damping of the dispersion energy. However, for noble-gas dimers the damping of the dispersion energy is not essential, and interatomic vdW potential can be effectively described even without a damping function, 
as was shown above. This provides additional reasoning why the scaling law for vdW radius~(\ref{eq.: Rvdw}), which was originally derived without considering dispersion damping~\cite{Fedorov2018}, works so well. 
To check the effect of damping function on the results, we derived the damped vdW-QDO potential, where the QDO damping function reads
\begin{equation}
    f_{2n}(z) = 1 - e^{-z} \sum_{k = 0}^{n} \frac{z^{k}}{k!}  \ , \ \ z = \frac{(\gamma R)^2}{2} \ .
\label{eq.: QDOdamp}
\end{equation}
It was found that for noble-gas dimers the obtained damped vdW-QDO potential curves are practically the same as the undamped ones within the considered range of distances (see Fig.~S2 in the Supporting Information). Interestingly, the damping function~(\ref{eq.: QDOdamp}) differs from the Tang-Toennies damping function~\cite{TangToennies1984} just in the upper summation limit (for the TT it is $2n$) and in the physical meaning of the unitless variable ($z = bR$ for the TT damping function, with $b$ stemming from the Born-Mayer repulsion term $Ae^{-bR}$). Note that, due to the distinct form of the Pauli repulsion in the vdW-QDO and TT models, the QDO damping function contains only even powers of $R$ up to $2n$ (see the derivation of the QDO damping function and more detailed discussion in the Supporting Infomation). 

We also observe that for both the TT and vdW-QDO models exchange repulsion and dispersion terms separately do not agree with their SAPT counterparts, whereas the total potentials show very close agreement with the sum of the SAPT terms (see the Supporting Information). This finding is in line with the statement of Ref.~\citenum{TangToennies1998} that the generalized Heitler-London theory delivers a more compact expansion of interaction energy than the SAPT. 

\subsection{Application to group II dimers}

Another class of vdW systems where Tang-Toennies potentials work well consists of Me$_2$ dimers, with Me = Mg, Ca, Sr, Ba, Zn, Cd, Hg. Although such systems are not purely vdW-bonded, it was demonstrated that their interatomic potentials can be also well described by the TT potential~\cite{TangToennies2008Hg, TangToennies2009Ca, TangToennies2010Mg, TangToennies2010Sr, TangToennies2011Ba, TangToennies2013Zn, TangToennies2015}. Moreover, potentials of the group II dimers also obey the principle of corresponding states, albeit the underlying potential shape is distinct from that of noble-gas dimers~\cite{TangToennies2010Sr,TangToennies2011Ba,TangToennies2013Zn}.
The only exception is the Be-Be dimer, which has been being a long-standing puzzle for quantum chemistry. The potential curve of Be$_2$ has a remarkably different shape in the long-range region, compared to other alkaline-earth elements~\cite{TangToennies2011Ba}. Since this dimer does not obey the principle of corresponding states, it is excluded from our consideration here. In what follows, we show that the vdW-QDO potential is also capable to describe the potential curves of the group II dimers upon several modifications. 

\begin{table}[h!]
\footnotesize
\caption{\footnotesize The reference \emph{ab initio} values of the dipole polarizability $\alpha_1$ (in a.u.), dipolar dispersion coefficient $C_6$ (in a.u.), and dimer potential well parameters $R_e$ (in bohr) and $D_e$ (in meV) of the group II dimers. For $R_e$ and $D_e$, the values in \AA~and Kelvin, respectively, are extra given in parentheses. The used data sources are the following: $^{\rm a}$\,Ref.~\citenum{Porsev2006}, $^{\rm b}$\,Ref.~\citenum{Ladjimi2023}, $^{\rm c}$\,Ref.~\citenum{AlphaDatabase2018}, $^{\rm d}$\,Ref.~\citenum{TangToennies2013Zn}, $^{\rm e}$\,Ref.~\citenum{Zaremba2021}, $^{\rm f}$\,Ref.~\citenum{Pahl2010}.}
\label{tab:metref}
\begin{tabular}{|c|c|c|c|c|}
\hline
~~ & ~$\alpha_1$~ & $C_6$ & $R_e$ (in \AA) & ~$D_e$ (in K)~ \\
\hline
~~Mg$_2$~~ & ~~71.3$^{\rm a}$~~ & ~~627$^{\rm a}$~~ & ~~7.35 (3.89)$^{\rm b}$~~ & ~~53.81 (624.4)$^{\rm b}$~~ \\
 Ca$_2$ & ~~157.1$^{\rm a}$~~ & ~2121$^{\rm a}$~ & ~8.13 (4.30)$^{\rm b}$~ & ~130.18 (1511)$^{\rm b}$~~ \\ 
 Sr$_2$ & ~197.2$^{\rm a}$~ & ~3103$^{\rm a}$~ & ~8.88 (4.70)$^{\rm b}$~ & ~129.69 (1505)$^{\rm b}$~~ \\ 
 Ba$_2$ & ~273.5$^{\rm a}$~ & ~5160$^{\rm a}$~ & ~9.43 (4.99)$^{\rm b}$~ & ~169.36 (1965)$^{\rm b}$~~ \\ 
 Zn$_2$ & ~38.67$^{\rm c}$~ & ~~359$^{\rm d}$~~ & ~7.23 (3.83)$^{\rm e}$~ & ~~28.64 (332.4)$^{\rm e}$~~ \\ 
 Cd$_2$ & ~~46$^{\rm c}$~~ & ~~686$^{\rm d}$~~ & ~7.32 (3.87)$^{\rm e}$~ & ~~40.91 (474.8)$^{\rm e}$~~ \\ 
 Hg$_2$ & ~~33.9$^{\rm c}$~~ & ~~392$^{\rm d}$~~ & ~6.95 (3.68)$^{\rm f}$~ & ~~48.60 (564.0)$^{\rm f}$~~ \\ 
\hline
\end{tabular}
\end{table}

First, in contrast to noble gases, for Me$_2$ dimers the damping 
function~(\ref{eq.: QDOdamp}) cannot be omitted due to much larger polarizabilities and hence dispersion coefficients than those of noble gases (see Table~\ref{tab:metref}). As a result, without damping function 
a pronounced divergence of the vdW-QDO potential would already occur at near-equilibrium distances. Second, the scaling laws of Eqs.~(\ref{eq.: eqdist}) and~(\ref{eq.: Descale}) are not so accurate for the group II dimers, since the bonding in Me$_2$ is not purely of vdW type~\cite{TangToennies2010Sr,TangToennies2011Ba}. Therefore, the reference values of $R_e$ and $D_e$ (Table~\ref{tab:metref}) were used for our vdW-QDO potential.

Following Tang \emph{et al.}~\cite{TangToennies2011Ba}, we choose Sr$_2$ dimer as the reference system to get the shape of the potential curve and then rescale it onto other dimers. Similar to the case of noble gases, the vdW-QDO potential shape for Sr$_2$ dimer was derived as (see the Supporting Information)
\begin{equation}
   U_{\rm QDO}^{\rm Sr}(x) =  \frac{A_d^*}{x} e^{-\frac{(\gamma^* x)^2}{2}} - \sum_{n=3}^{5} f_{2n}(\gamma^* x)\frac{C_{2n}^*}{x^{2n}}\ ,
\label{eq.: U_Sr2}
\end{equation}
with the numerical values of its parameters presented in Table~\ref{tab:metpotpar}. The QDO parameters $\{ q, \mu, \omega \}$ for the Sr$_2$ dimer were set using $\alpha_1, C_6$ and $R_e$ following the damped vdW-OQDO procedure, as explained in the Supporting Information. Altogether, the three physical quantities are employed to parametrize the vdW-QDO potential for Sr$_2$, compared to the five $\{ R_e, D_e, C_6, C_8, C_{10} \}$ in case of the Tang-Toennies potential~\cite{TangToennies2011Ba}. For other Me$_2$ dimers, we need only $R_e$ and $D_e$ from Table~\ref{tab:metref} to perform the rescaling
\begin{equation}
 V_{\rm QDO}(R) = D_e U_{\rm QDO}^{\rm Sr}(x)\ , \ \ x = {R}/{R_e}\ .
\label{eq.: Vmet}
\end{equation}

\begin{table}[h!]
\footnotesize
\caption{\footnotesize The dimensionless parameters in Eq.~(\ref{eq.: U_Sr2}). The strontium dimer parameters used in the second column are $D_e({\rm Sr}_2) = 129.7~{\rm meV} = 4.766 \times 10^{-3}~{\rm a.u.}$ and $R_e({\rm Sr}_2) = 8.88$ bohr~\cite{Ladjimi2023}. The QDO parameters for Sr$_2$ dimer are $q = 1.5433$, $\mu = 1.0671$, $\omega = 0.1064$ (all in a.u.).}
\label{tab:metpotpar}
\begin{center}
\begin{tabular}{|c|c|c|}
\hline
~~Parameter~~ & ~Definition~ & ~~Numerical value~~ \\
\hline
~$A_d^*$~ & ~$A_d k_e q^2/R_e D_e$~ & ~58.051~ \\
~$\gamma^*$~ & ~$R_e\sqrt{\mu\omega/\hbar}$~ & ~2.992~ \\
~$C_6^*$~ & ~$C_6/D_eR_e^6$~ & ~1.6209~ \\
~$C_8^*$~ & ~$5C_6/D_eR_e^6(\gamma^*)^2$~ & ~0.9053~ \\
~$C_{10}^*$~ & ~~$245C_6/8D_eR_e^6(\gamma^*)^4$~~ & ~0.6194~ \\
\hline
\end{tabular}
\end{center}
\end{table}

\begin{figure*}[t!]
\includegraphics[width=1.2\LL]{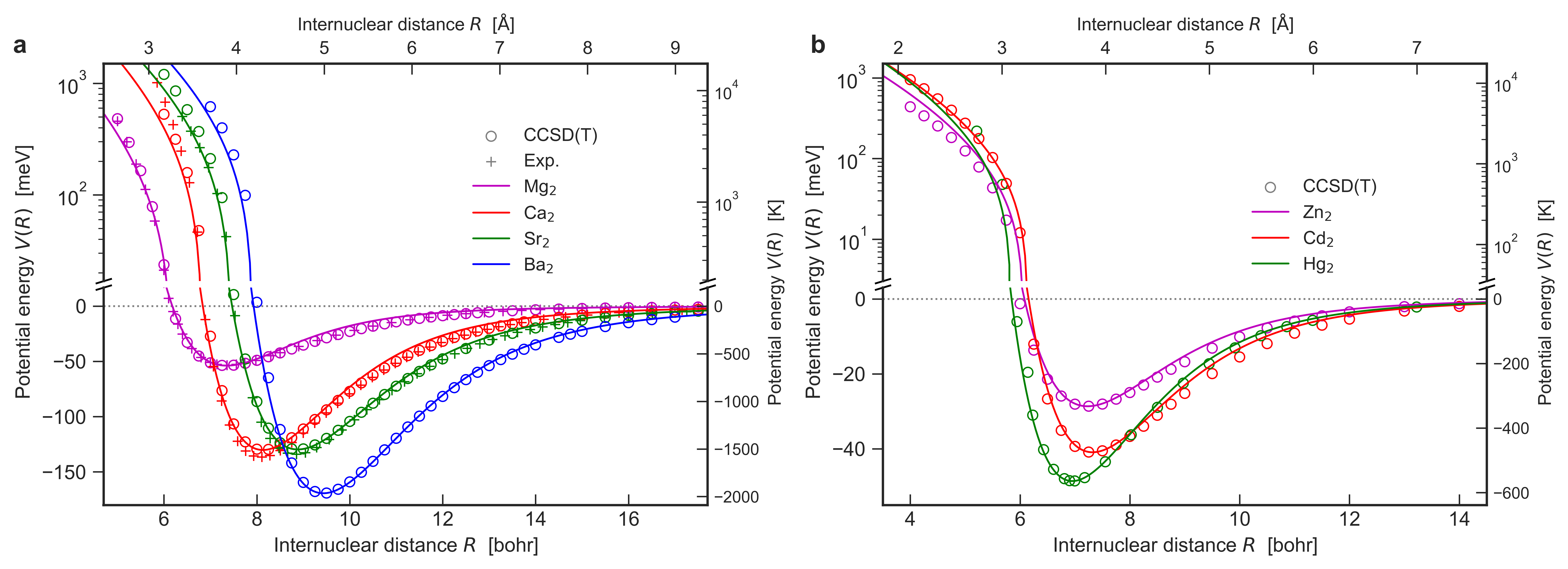}
\caption{\footnotesize Interatomic potentials of (a) Mg$_2$, Ca$_2$, Sr$_2$, Ba$_2$ and of (b) Zn$_2$, Cd$_2$, Hg$_2$ dimers. The vdW-QDO potentials are shown by solid lines, circles mark coupled-cluster calculations~\cite{Ladjimi2023,Zaremba2021,Pahl2010}, and crosses display experimental potential curves~\cite{Tiesinga2002, Allard2002, Stein2008}.}
\label{fig.: metals}
\end{figure*}

\begin{figure}[h!]
\includegraphics[width=0.7\LL]{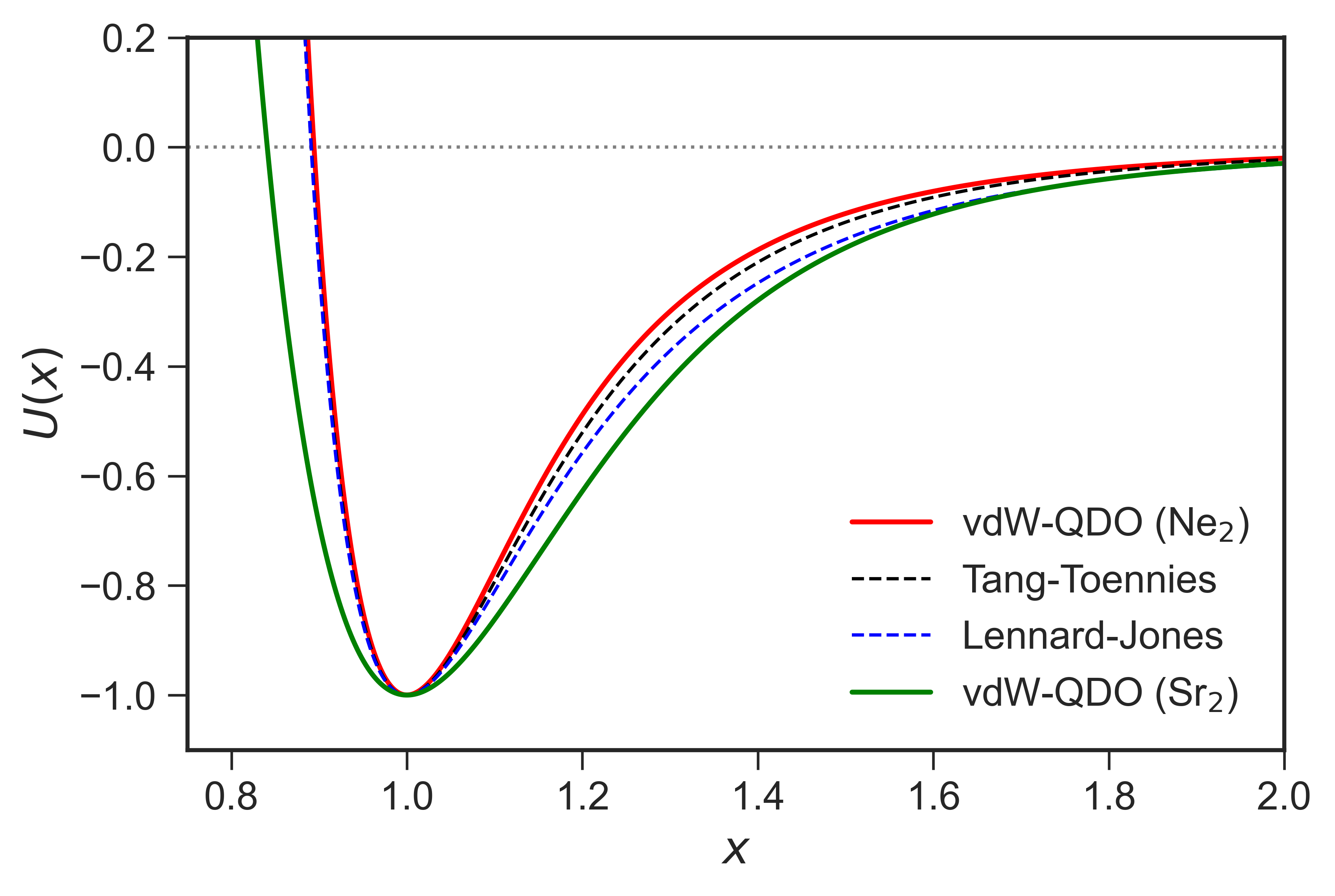}
\caption{\footnotesize Dimensionless shapes of the vdW-QDO potential curves for Ne$_2$ (red) and Sr$_2$ (green) compared to the shapes of the LJ (dashed blue) and TTS (dashed black) potentials.}
\label{fig.: potshapes}
\end{figure}

The results in Fig.~\ref{fig.: metals} show that vdW-QDO potentials are in excellent agreement with both \emph{ab initio} and experimentally derived potentials (when they are available). This is a remarkable result, since the binding energies of group II dimers are up to 5 times larger than those of the heaviest noble gases. Furthermore, the shape of their potentials is distinct to the one of noble gases (Fig.~\ref{fig.: potshapes}). Thus, the vdW-QDO functional form is robust and well-suited to describe vdW potentials across various types of systems. In contrast, the LJ potential cannot be employed to describe group II dimers with any combination of parameters, since its energy well is too narrow for such binding curves (Fig.~\ref{fig.: potshapes}).

\subsection{Application to molecular dimers}

Finally, we can show that the developed vdW-QDO potential is also applicable to molecular systems, with an example of eight dispersion-dominated molecular dimers from the S66$\times$8 benchmark dataset~\cite{Rezac2011}.
They are hydrocarbons, including benzene as well as aliphatic and cyclic molecules (see the list in the Supporting Information). Such systems were chosen to diminish an influence of electrostatic term not included in the current vdW-QDO potential.

We compute the energy of vdW interaction between molecules $A$ and $B$ at given intermolecular separation as
\begin{equation}
 V^{\rm vdW}_{\rm QDO}(A,B) = V^{\rm exc}_{\rm QDO} + V^{\rm disp}_{\rm QDO} = \sum_{i \in A} \sum_{j \in B} V_{\rm QDO}^{ij}(R_{ij})\ ,
\label{eq.: Vtotmol}
\end{equation}
where summation goes over the atoms $i$ and $j$ of the molecules $A$ and $B$, respectively, and $R_{ij}$ is the interatomic distance. Interaction energy between a pair $(i,j)$ is given by the damped vdW-QDO potential
\begin{equation}
   V_{\rm QDO}^{ij}(R_{ij}) = A_d \frac{k_e q^2}{R_{ij}} e^{-\frac{(\gamma R_{ij})^2}{2}} - \sum_{n=3}^{5} f_{2n}(\gamma R_{ij})\frac{C_{2n}}{R_{ij}^{2n}}\ .
\label{eq.: Vmol}
\end{equation}
To set the QDO parameters $\{ q, \mu, \omega \}$, we apply
the vdW-OQDO procedure (see the Supporting Information) coupled with the atom-in-molecule (AIM) approach to each pair of atoms $(i,j)$. Following the Tkatchenko-Scheffler (TS) method~\cite{Tkatchenko2009}, the AIM polarizabilities and dispersion coefficients are defined by
\begin{equation}
   \alpha_{1,i}^{\rm AIM} = \alpha_{1,i}^{\rm free} \left( \frac{V_i^{\rm AIM}}{V_i^{\rm free}} \right), \ C_{6,i}^{\rm AIM} = C_{6,i}^{\rm free} \left( \frac{V_i^{\rm AIM}}{V_i^{\rm free}} \right)^2,
\label{eq.: AIMparam}
\end{equation}
where $V_i^{\rm free}$ and $V_i^{\rm AIM}$ are the corresponding Hirshfeld volumes. To compute them, single-point DFT-PBE0~\cite{Adamo1999} calculations for every dimer were performed at their equilibrium geometry. Then, the effective polarizability $\alpha_1^{ij}$ and dispersion coefficient $C_6^{ij}$ for a pair $(i,j)$ were defined using the combination rules of Eqs.~(\ref{eq.: c6mix}) and (\ref{eq.: alphamix}). Finally, the vdW-OQDO parametrization procedure (see the Supporting Information) was applied to map $\{ \alpha_1^{ij}, C_6^{ij} \}$ onto $\{ q, \mu, \omega \}$. After performing pairwise summation in Eq.~(\ref{eq.: Vtotmol}) and repeating the whole procedure for all 8 intermolecular separations, the vdW-QDO interaction curves for dimers are obtained and compared to the reference CCSD(T) interaction curves~\cite{Rezac2011}. For comparison, interaction energies of dimers at PBE0+TS~\cite{Tkatchenko2009}, PBE0+MBD~\cite{Tkatchenko2012}, and DFTB3~\cite{dftb2020}+MBD levels of theory were also calculated. All DFT calculations in this work were performed using FHI-aims code~\cite{Blum2009} with the \lq tight\rq\ atomic basis sets.

\begin{figure*}[t!]
\includegraphics[width=1.2\LL]{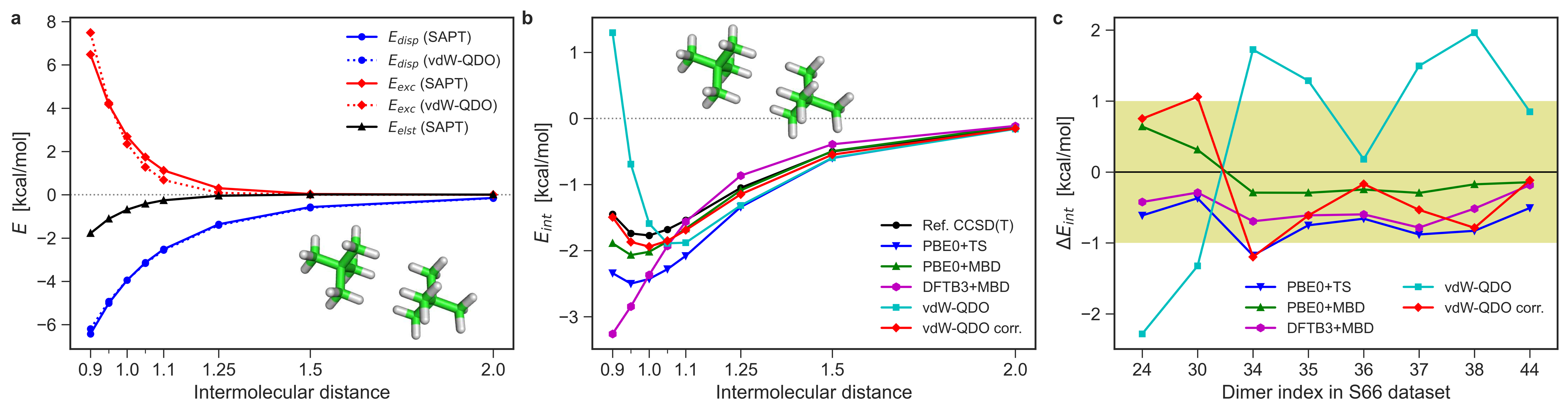}
\caption{\footnotesize (a) Dispersion (blue) and exchange (red) contributions to the interaction energy of neopentane dimer (shown as inset) calculated by SAPT-DFT~\cite{Hesselmann2018} (solid lines) and damped vdW-QDO potential of Eq.~(\ref{eq.: Vtotmol}) (dotted lines). In addition, electrostatic term from SAPT-DFT is displayed in black. (b) Binding energy curves of neopentane dimer as calculated by different methods: CCSD(T)~\cite{Rezac2011} (black); PBE0+TS (blue) and PBE0+MBD (green); DFTB3+MBD (magenta); damped vdW-QDO potential (cyan); SAPT-corrected vdW-QDO potential~(\ref{eq.: Vmolsapt}) (red). (c) Errors in interaction energy of eight dispersion-dominated dimers for the five methods considered. Yellow filling depicts the ``chemical accuracy'' region of $\pm1$ kcal/mol error.}
\label{fig.: molecules}
\end{figure*}

We showcase the obtained results with an example of neopentane dimer (Fig.~\ref{fig.: molecules}b). One can see that PBE0+TS overbinds neopentane dimer significantly and underestimates the equilibrium separation by 5\%. Inclusion of many-body effects at the PBE0+MBD level improves the results of PBE0+TS for energy, but the predicted equilibrium separation is still 95\% of the reference value. On the other hand, DFTB3+MBD method clearly lacks repulsion and attraction at short-range and long-range distances, respectively, although at the equilibrium distance the two errors largely cancel each other. As for the vdW-QDO potential, we note that the minimum of intermolecular interaction curve is very close to the reference CCSD(T) energy, although our method overestimates the equilibrium separation by 10\%. When considering the overall interaction curve, however, the vdW-QDO potential deviates significantly from the reference, being too repulsive at shorter distances. These conclusions remain valid also for other seven dimers, as supported by the corresponding results in the Supporting Information.

To get insights into such a behavior, we compared the exchange and dispersion terms in vdW-QDO potential~(\ref{eq.: Vtotmol}) to their counterparts from SAPT-DFT calculations~\cite{Hesselmann2018} (see Fig.~\ref{fig.: molecules}a). The dispersion part of vdW-QDO potential provides excellent agreement with $E_{\rm disp}^{(2)}$ from SAPT-DFT, 
which illustrates the well-known fact that vdW dispersion interaction between small molecules can be effectively described by a pairwise potential, despite being essentially many-body in its nature~\cite{Tkatchenko2012,Hermann2017-Chemrev,Hermann2017}. In contrast, for the vdW-QDO exchange repulsion term we observe a noticeable deviation from the $E_{\rm exc}^{(1)}$ contribution of SAPT-DFT. This might indicate that exchange repulsion between molecules is not accurately described by the commonly used pairwise approach and hence exchange repulsion requires many-body treatment as well~\cite{Stoehr2020}.

SAPT-DFT calculations~\cite{Hesselmann2018} also indicate that electrostatic contributions are small but not negligible even for dispersion-dominated dimers (Fig.~\ref{fig.: molecules}a and Figs.~S6--S8 in the Supporting Information). To check a possible effect of inclusion of accurate many-body exchange and electrostatic interactions on our results, we consider the corrected vdW-QDO potential, where the first-order SAPT-DFT energy was added to the dispersion energy from the vdW-QDO model
\begin{equation}
  V^{\rm corr}_{\rm QDO}(A,B) = E_{\rm elst}^{(1)} + E_{\rm exc}^{(1)} + V^{\rm disp}_{\rm QDO}\ .
\label{eq.: Vmolsapt}
\end{equation}
This approach is similar to the HFDc$^{(1)}$ scheme of Podeszwa \emph{et al.}~\cite{Podeszwa2010} with the important difference that here the dispersion energy is calculated from the QDO model, whereas in the HFDc$^{(1)}$ method this energy is computed from SAPT. The corrected vdW-QDO curve in Fig.~\ref{fig.: molecules}b delivers much better description of the interaction energy than the original vdW-QDO method. Notably, at larger distances $V^{\rm corr}_{\rm QDO}$ also shows a good agreement with PBE0+MBD energy. 

The overall statistics for the eight molecular dimers is shown in Fig.~\ref{fig.: molecules}c in terms of the error in the equilibrium energy obtained by five methods with respect to the reference CCSD(T) values~\cite{Rezac2011}. Purely analytical vdW-QDO predictions are scattered within 2 kcal/mol range, which is remarkable considering the approximations made in the model. Including the first-order SAPT energy reduces errors roughly to 1 kcal/mol (chemical accuracy).

This test demonstrates that the vdW-QDO approach is not limited to atomic dimers and can be generalized to molecular complexes, although in that case the accuracy is currently lower. Nevertheless, considering the approximations made, the fact that the vdW-QDO method (even without SAPT corrections) properly predicts binding of the considered weakly-bound molecular complexes from a set of AIM quantities ($\alpha_1$ and $C_6$) is already reassuring.

\section{Summary and outlook}

We developed a universal pairwise vdW potential devoid of empiricism and parameterized by only two atomic non-bonded parameters. The developed vdW-QDO potential combines the strengths of the Lennard-Jones and Tang-Toennies models. Similarly to the former, the vdW-QDO potential is fully determined by only two parameters. At the same time, our potential is comparable in accuracy to the Tang-Toennies potential for noble-gas dimers, being twice as accurate as the Lennard-Jones one. Moreover, the vdW-QDO potential has advantages which are present neither in Lennard-Jones nor in Tang-Toennies models. First, the two parameters $\{ \alpha_1, C_6 \}$ are readily available for the whole periodic table~\cite{Gobre2016,AlphaDatabase2018}, being computed by highly accurate \emph{ab initio} methods. This makes our potential widely applicable, as demonstrated for atomic dimers of group II elements as well as organic molecular dimers. Second, the conformal vdW-QDO potential has better long-range behavior than the Lennard-Jones and conformal Tang-Toennies potentials. This is crucial for applications to extended systems, where errors in the long-range vdW energy accumulate over many atomic pairs.

The key idea behind the presented potential is the synergy between vdW scaling laws, coarse-grained QDO formalism for exchange repulsion, and the principle of corresponding states. In its current form, the vdW-QDO potential does not explicitly include the electrostatic contribution arising from charge penetration between atoms at short distances, which is non-negligible according to the SAPT. Although the short-range electrostatic interaction can be introduced into the QDO model in the form of Coulomb integral~\cite{Sadhukhan2016}, this requires introduction of an additional ``electrostatic charge'' parameter~\cite{Martyna2019-RevModPhys} into the vdW-QDO potential. Therefore, given the good accuracy of the current vdW-QDO potential, we decided to dispense with the explicit modeling of short-range electrostatics at this stage of the model development. In its current version, the vdW-QDO potential can be incorporated into classical force fields, as a non-empirical replacement for the LJ potential. For polar systems, an additional electrostatic/polarization term is needed in a force field (like it is done in case of the LJ potential) to complete the description of non-covalent interactions.

We consider the present vdW-QDO potential as an important step towards a new generation of universal vdW potentials to be used in force fields and for biomolecular applications. To extract atom-in-molecule parameters, we currently employ \emph{ab initio} calculations, which is a certain limitation. However, active development of machine-learning models~\cite{Muhli2021, Westermayr2022, Piquemal2022-MBD2} paves the way to obtain atom-in-molecule partitioning without costly electronic-structure calculations. Moreover, there is an increasing trend in creating extensive molecular datasets such as QM7-X~\cite{Hoja2021QM7-X}, which include information about atom-in-molecule volumes. This enables the direct applicability of the vdW-QDO potential to arbitrary molecular systems. Finally, the vdW-QDO potential can be generalized to include polarization and electrostatic contributions, which is the subject of our ongoing studies. Eventually, this effort should deliver a general coarse-grained force field for all non-covalent interactions entirely based on the model system of coupled QDOs.

\begin{acknowledgement}
We acknowledge financial support from the Luxembourg National Research Fund \emph{via} FNR ``ACTIVE (PRIDE19/14063202)'' and ``QuantPhonon (INTER/DFG/18/12944860)'' projects as well as from the European Research Council (ERC Advanced Grant ``FITMOL''). We thank Prof. Micha{\l} Tomza for providing us with the numerical data for the coupled-cluster potential curves of alkaline-earth dimers before their publication. We also thank Dr. Leonardo Medrano Sandonas (Uni.lu) for sharing his DFTB3+MBD results on molecular dimers.
\end{acknowledgement}

\begin{suppinfo}
%
%A listing of the contents of each file supplied as Supporting Information should be included. For instructions on what should be included in the Supporting Information as well as how to prepare this material for publications, refer to the journal's Instructions for Authors.
%
The Supporting Information is available free of charge at ...
\begin{itemize}
  \item vdW-OQDO parameters for atomic dimers, derivation of the QDO damping function and the damped vdW-QDO potential, the results for eight molecular dimers.
%  \item Filename: brief description
\end{itemize}
\end{suppinfo}

\bibliography{biblio}

\end{document}